\documentclass{article}

\usepackage{arxiv}

\usepackage[utf8]{inputenc} 
\usepackage[T1]{fontenc}    
\usepackage{hyperref}       
\usepackage{url}            
\usepackage{booktabs}       
\usepackage{amsfonts}       
\usepackage{nicefrac}       
\usepackage{microtype}      
\usepackage{lipsum}
\usepackage{graphicx}
\usepackage{bm}
\usepackage{amsmath, amssymb}
\usepackage{mathptmx}
\usepackage{mathrsfs}

\newcommand{\koop}{\mathcal{K}_{{\it{F}}}}
\newcommand{\koopm}{\bm{K}_{{\it{F}}}}

\newcommand{\F}{{\it{F}}}
\newcommand{\x}[1]{\bm{x}_{#1}}
\newcommand{\argmax}{\mathop{\rm arg~max}\limits}

\title{Hybrid scheme of kinematic analysis \\ and   Lagrangian Koopman operator analysis \\ for short-term precipitation forecasting}

\author{
 Shitao Zheng \\
  Graduate School of Engineering\\
  University of Yamanashi\\
  Kofu, Yamanashi, Japan\\
  \texttt{g18tca02@yamanashi.ac.jp} \\
   \And
 Takashi Miyamoto\thanks{Corresponding Author} \\
  Department of Civil and Environmental Engineering\\
  University of Yamanashi\\
  Kofu, Yamanashi, Japan\\
  \texttt{tmiyamoto@yamanashi.ac.jp}
   \And
 Koyuru Iwanami \\
  Storm, Flood and Landslide Research Division\\
  National Research Institute for 
  Earth Science and Disaster Resilience (NIED)\\
  Tsukuba, Ibaraki, Japan\\
  \texttt{iwanami@bosai.go.jp}
     \And
 Shingo Shimizu \\
  Storm, Flood and Landslide Research Division\\
  National Research Institute for 
  Earth Science and Disaster Resilience (NIED)\\
  Tsukuba, Ibaraki, Japan\\
  \texttt{shimizus@bosai.go.jp}
     \And
 Ryohei Kato \\
  Storm, Flood and Landslide Research Division\\
  National Research Institute for 
  Earth Science and Disaster Resilience (NIED)\\
  Tsukuba, Ibaraki, Japan\\
  \texttt{rkato@bosai.go.jp}
}

\begin{document}
\maketitle
\begin{abstract}
With the accumulation of meteorological big data, data-driven models for short-term precipitation forecasting have shown increasing promise. We focus on Koopman operator analysis, which is a data-driven scheme to discover governing laws in observed data. We propose a method to apply this scheme to phenomena accompanying advection currents such as precipitation. The proposed method decomposes time evolutions of the phenomena between advection currents under a velocity field and changes in physical quantities under Lagrangian coordinates. The advection currents are estimated by kinematic analysis, and the changes in physical quantities are estimated by Koopman operator analysis. The proposed method is applied to actual precipitation distribution data, and the results show that the development and decay of precipitation are properly captured relative to conventional methods and that stable predictions over long periods are possible.
\end{abstract}


\section{Introduction}
The rise of data science accompanying the accumulation of big data in recent years marks an important paradigm shift. The methods of data science, represented by machine learning, not only advance the automation and precision of various kinds of task processing through the development and performance improvement of numerical models but also create a new analysis framework for phenomena whose complexity previously made them difficult to analyze.

Precipitation is a natural phenomenon associated with physical processes at various scales, from atmospheric circulation to the behavior of aerosol particles. The accurate prediction of precipitation has been an important theme in meteorology and serves an important role in the prevention of meteorological hazards, which have increased in recent years. Precipitation forecasting is generally conducted using the numerical analysis of fundamental governing equations that describe various physical phenomena. However, prediction accuracy, even minutes to hours ahead of time, is hampered by factors of numerical models such as the spatial averaging of microscale physical phenomena and of numerical analysis relating to the spatial discretization or initial value settings of the equations used. These are major issues facing the reliability of short-term predictions.

In this context, data-driven precipitation forecasting models that use meteorological big data have shown increasing promise. There has been extensive research on applying machine learning methods such as deep learning, which saw significant developments in precipitation prediction performance in the 2010s. However, despite the proposals of specific computational models, no breakthroughs have been achieved.

One reason why conventional data-driven approaches have not succeeded in precipitation forecasting is that it is difficult to apply machine learning frameworks, which are essentially interpolation techniques for learning data, to the temporal extrapolation of complex phenomena whose patterns are difficult to understand from observed data alone. 

With this in mind, the present research focuses on new trends of data science that seek to design computational models with acquired extrapolation capabilities by discovering the laws and fundamental equations underlying these phenomena in a data-driven manner. This manuscript focuses on Koopman operator analysis, which is one such computational model, and proposes a novel method for applying this framework to phenomena accompanying advection currents in order to create a precipitation forecasting model based on this analysis method. The proposed method decomposes time evolutions of the phenomena between advection currents of physical quantities under a velocity field and changes in physical quantities under Lagrangian coordinates. Kinematic analysis is used to analyze and predict the advection currents, and Koopman operator analysis is used to determine the physical quantities, making this a data-driven model based on dynamic aspects of the phenomena.

The manuscript is structured as follows. Chapter 2 organizes the related works into two perspectives—data-driven approaches relating to precipitation phenomena and data-driven science attempting to discover fundamental laws and equations—and discusses the novelty of the present research compared to these viewpoints. Chapter 3 provides an overview of Koopman operator analysis and proposes an analysis method for phenomena accompanying advection currents, where the conventional application of similar methods was difficult to apply. Chapter 4 conducts short-term predictions of precipitation based on the proposed method using actual precipitation data, and the effectiveness is verified by comparing the results with conventional methods. Chapter 5 presents a discussion of the verification results, and Chapter 6 discusses the findings from the present research and its future potential.

\section{Related works}
\subsection{Data-driven approach for precipitation forecasting}
The principal fundamental equation systems relating to precipitation phenomena can be broadly classified between those relating to cloud dynamics processes, which describe the fluid dynamics of the atmosphere \cite{holton}, and those relating to cloud microphysics, which describe the behavior of cloud particles and precipitation particles \cite{prup}. The prediction of precipitation is generally determined by numerically solving these equations, but the scale of the cloud microphysics is small relative to the spatial scale of the analysis domain and is difficult to directly express. Therefore, various types of spatial averaging and empirical formulations have been conducted \cite{khain}, and accuracy of these models is still being improved. Furthermore, it is well known that the prediction accuracy of the initial analysis period of meteorological phenomena has been hampered by spin-up problems such as unnatural behavior in the initial analysis stage caused by physical imbalances in the initial values of the model \cite{zhao}.

Researchers have begun to develop precipitation prediction models based on data-driven methods to overcome the abovementioned issues faced in numerical analysis. Shi et al. proposed a deep learning model structure that learns time-series data of the precipitation distributions obtained from radar observations and applied it to precipitation prediction \cite{shi1, shi2}. Agrawal et al. proposed a deep learning method that takes observed precipitation amounts obtained from multiple multispectral radar sites as an input \cite{google}.

Although deep learning applications have become more mainstream in the development of precipitation prediction models based on data-driven methods, quantitative precipitation prediction has not yet seen sufficient degrees of success. An underlying cause is that deep learning techniques are fundamentally pattern recognition methods that interpolate learning data in a feature space and cannot readily comprehend the patterns generated from complex natural phenomena from past data alone. It is thought that deep learning models operate under the manifold hypothesis \cite{manifold}, where most real-world data exist as low-dimensional manifolds wherein inter-data feature differences are reflected in their distance structure. By learning these manifold structures, deep learning models are thought to obtain generalization performance against the unknown data as manifold interpolations \cite{repre}. However, it is thought that a computational model that can conduct extrapolations to areas not in the learning data is necessary for predicting the time evolutions of complex phenomena.

\subsection{Data-driven discovery of fundamental laws and Koopman operator analysis}
Recently, researchers have attempted to discover not only the behavior of the data itself but also the fundamental laws followed by the data in a data-driven approach. Models that express the fundamental laws followed by the data are anticipated to achieve high extrapolation performance by conducting computations based on laws that are applicable in various situations.

Assuming that the time evolutions laws of the output variables are a linear sum of the nonlinear functions of the input variables, Brunton et al. proposed a method that identified predominant terms of governing equations using sparse regression \cite{sindy, pdefind}. Raissi et al. assumed that the solution of the governing equations was a Gaussian process and developed methods that determined parameters in equations with maximum likelihood estimation \cite{raissi1, raissi2}. Berg et al. used a neural network to express the time evolution laws followed by the variables of interest and proposed a method to obtain a specific model using backpropagation \cite{berg}.

The Koopman operator analysis method has added to these advances by broadly discovering nonlinear time evolution laws in a data-driven manner with fewer assumptions \cite{kmd-org}. Based on the operator analysis of arbitrary functions that act on the data, Koopman operator analysis expresses the nonlinear time evolution laws in the form of a linear mode decomposition and has been applied in the analysis of fluid phenomena \cite{kmd, mezic}. The Koopman operator analysis method is a data-driven model that has favorable characteristics in the analysis of natural phenomena, given its ability to assume arbitrary nonlinear laws underlying the computational model and explicitly reflect prior knowledge relating to the phenomena into the computational model \cite{kmdpde}.

In principle, it is difficult to apply Koopman operator analysis to phenomena accompanying advection currents owing to its mathematical characteristics. The present research focuses on this issue in applying Koopman operator analysis to precipitation forecasting. The novelty of the present research lies in applying Koopman operator analysis to physical changes in the phenomena under Lagrangian coordinates, proposing a specific computational method, and verifying its effectiveness as applied to actual observed data.

\section{Methods}
\subsection{Koopman operator analysis}
Koopman operator analysis is a method for reconstructing the characteristics of nonlinear time evolution laws from observed data based on mathematical theories that transform nonlinear time evolution laws of dynamic phenomena to mode expansions of linear laws. The following presents an overview of the mode decomposition method based on a Koopman operator analysis of the discrete system and its characteristics. 

We consider a state variable $\x \in \it{M}$ on a manifold $\it M$, 
discretely transitioning to the next time step under a nonlinear mapping $\F$, as shown in the equation below.
\begin{equation}
    \x{t+1} = \F (\x{t})
    \label{eqn:statespace}
\end{equation}
Given the functions $g_i:{\it{M}} \rightarrow \bm{R} (i=1,\cdots, N)$ in the Hilbert space, referred to as the observable functions, the linear operator $\koop$, referred to as the Koopman operator, acting on $g_i$ is defined as shown in the following equation \cite{koopman}.
\begin{equation}
    \koop g_i \equiv g_i \circ \F
    \label{eqn:def_koop}
\end{equation}
The following equation on the variable $\bm{y}=( g_1(\bm{x}) \cdots g_N(\bm{x}) )$ due to $\bm{g}=(g_1 \cdots g_N)$ holds using spectrum $\lambda_j$ and eigenfunction $\phi_j(j=1,\cdots,\infty)$ of $\koop$.
\begin{eqnarray}
    \bm{y}_t=\bm{g}(\bm{x}_t)=\koop \bm{g}(\bm{x}_{t-1})&=&\cdots=(\koop)^t \bm{g}(\bm{x}_0)=(\koop)^t \sum_{j=1}^{\infty} \phi_j(\bm{x}_0)\bm{v}_j
    = \sum_{j=1}^{\infty} (\lambda_j)^{t} \phi_j(\bm{x}_0)\bm{v}_j \nonumber \\
 \text{where} \
    \bm{v}_j &=& \begin{pmatrix}
                  v_{1j} &  ... & v_{Nj}
                  \end{pmatrix}
     \label{eqn:kmd}
\end{eqnarray}
where $v_{ij}$ is the expansion coefficient of $g_i$ due to $\phi_j$. The above equation indicates that the time evolution of $\bm{g}(\bm{x})$ can be decomposed as a sum of modes $\bm{v}_j$ with the time vibration $(\lambda_j)^t$. $\bm{v}_j$, $\lambda_j$, and Eq. (\ref{eqn:kmd}) are referred to as the Koopman mode, Koopman eigenvalue, and Koopman mode decomposition (KMD), respectively. The Koopman mode $\bm{v}_j$ expresses the spatial distribution patterns of dynamic phenomena following (\ref{eqn:statespace}), and the Koopman eigenvalue expresses the frequency and amplification/decay of the vibration of each spatial mode. 

In the manner shown above, the nonlinear law (\ref{eqn:statespace}) of the data series $\{\bm{x}_0,\cdots, \bm{x}_t \}$ of a given dynamic physical variable can be expressed in the form of a mode decomposition using the function g, as shown in (\ref{eqn:kmd}). The mode shape and eigenvalue can be estimated from the series $\{\bm{y}_0,\cdots, \bm{y}_t \}= \{\bm{g}(\bm{x}_0), \cdots, \bm{g}(\bm{x}_t) \}$ using algorithms such as dynamic mode decomposition \cite{tu}, and the time evolution of the data following (\ref{eqn:statespace}) can be obtained by temporally extrapolating the obtained modes, without solving the original equations.

When applying this method to actual data, the infinite series in (\ref{eqn:kmd}) must be converted to the finite series using the restriction $\koopm$ on the subspace of the Koopman operator $\koop$, assuming that each $g_i$ belongs to an $M$-dimensional subspace that is invariant to the Koopman operator (Koopman invariant subspace). 
\begin{eqnarray}
    \bm{y}_t=(\koopm)^t \bm{g}(\bm{x}_0)
    = \sum_{j=1}^{M} (\lambda_j)^{t} \phi_j(\bm{x}_0)\bm{v}_j
     \label{eqn:kmd2}
\end{eqnarray}
Generally, the $g_i$ belonging to the Koopman invariant subspace is determined from prior knowledge relating to the original law (\ref{eqn:statespace}) \cite{kmdpde}, and there have been attempts to construct $g_i$ in a data-driven manner via machine learning methods \cite{learning}. Koopman operator analysis is equivalent to standard dynamic mode decomposition when using an identity function as $g_i$, and the present research uses an identity function during computations for the sake of simplicity.

\subsection{Hybrid scheme of kinematic analysis and Koopman operator analysis}
As mentioned above, Koopman operator analysis analyzes data characteristics and predicts time evolutions based on mode decompositions of $\{\bm{y}_0,\cdots, \bm{y}_t\}=\{\bm{g}(\bm{x}_0), \cdots, \bm{g}(\bm{x}_t) \}$. The Koopman modes $\bm{v}_j$ obtained from the method express patterns that are spatially fixed, and a known problem is that mode decompositions cannot be properly conducted on phenomena whose physical quantities globally move with time accompanying advection currents \cite{cdmd}.

With this in mind, the present research decomposes temporal changes in physical quantities between global advection currents and physical changes under Lagrangian coordinates and proposes a method for estimating the advection currents with kinematic analysis and the changes in physical quantities with Koopman operator analysis.

We set $\bm{r}$ as the Eulerian coordinate of the transformed variable $\bm{y}$ of the physical variable $\bm{x}$ and assume that the data series $\{ \bm{y}_0(\bm{r})\cdots,\bm{y}_t(\bm{r}) \}$ is obtained. At this point, the data series coordinates are converted to Lagrangian coordinates $\bm{R}$, based on information on the velocity field $\bm{u}(\bm{r},t)$.
\begin{equation}
    \bm{y}_k(\bm{R}) =\bm{y}_k(\bm{R}(\bm{r}, \bm{u}, k))
\end{equation}
The mode in (\ref{eqn:kmd2}) is estimated by applying a dynamic mode decomposition on the data series $\{ \bm{y}_0(\bm{R})\cdots,\bm{y}_t(\bm{R}) \}$ obtained from this conversion, and the time evolution is predicted as $\{ \bm{y}_{t+1}(\bm{R}),\bm{y}_{t+2}(\bm{R})\cdots \}$ from temporal extrapolation. The prediction series obtained in this way can be once again converted to Eulerian coordinates based on the predicted values of the velocity field $\bm{u}(\bm{r},t)$ to obtain the prediction series $\{ \bm{y}_{t+1}(\bm{r}),\bm{y}_{t+2}(\bm{r})\cdots\}$.

For the sake of simplicity, we approximate that the velocity field is not dependent on time or coordinates and is uniform.
\begin{equation}
    \bm{u}(\bm{r}, t) \simeq \bm{u}_0
    \label{eqn:uni}
\end{equation}
The relationship between the Eulerian coordinates $\bm{r}$ and Lagrangian coordinates $\bm{R}$ is shown below.
\begin{equation}
    \bm{r} = \bm{R} - \bm{u}_0 \cdot t
\end{equation}
When (\ref{eqn:uni}) holds as a good approximation, $\bm{u}_0$ can be kinematically determined from the data as a value that maximizes the cross-correlations of the physical quantities in two periods, as shown below.
\begin{eqnarray}
\bm{u}_0 &=& \argmax_{\bm{u}} \sum_{\bm{r}} \bm{x}_{t_1}(\bm{r}) \cdot \bm{x}_{t_2}(\bm{r} + \bm{u} \Delta t) \nonumber \\
\text{where}  \ \Delta t &=& t_2 - t_1
\label{eqn:derive_uni}
\end{eqnarray}

The above process is summarized as an algorithm that uses the observed data series $(\bm{x}_0(\bm{r}), \cdots, \bm{x}_n(\bm{r}))$ to determine its time evolution.

\paragraph{Algorithm}
\begin{enumerate}
    \item The velocity field $\bm{u}_0$ is determined
    from the data $\bm{x}_i(\bm{r}), \bm{x}_j(\bm{r})$ at two given time steps.
      \begin{eqnarray}
       \bm{u}_0 &=& \argmax_{\bm{u}} \sum_{\bm{r}} \bm{x}_i(\bm{r}) \cdot \bm{x}_j(\bm{r} + \bm{u} \Delta t) \nonumber \\
        \text{where} \ \Delta t &=& j - i \ \ (j > i) \nonumber
      \end{eqnarray}
    \item  $\bm{u}_0$ is used to convert from the physical variable distribution$\{ \bm{x}_0(\bm{r}), \cdots, \bm{x}_n(\bm{r}) \}$ in Eulerian coordinates to the distribution $\{ \bm{x}_0(\bm{R}), \cdots, \bm{x}_n(\bm{R}) \}$ in Lagrangian coordinates.
        \begin{eqnarray}
        \bm{x}_i(\bm{R}) = \bm{x}_i(\bm{r} + \bm{u}_0\cdot i) \ \ (i=0,\cdots, n)\nonumber
        \end{eqnarray}
    \item $\bm{y}_i(\bm{R}) = \bm{g}(\bm{x}_i(\bm{R})) (i=0,\cdots,n)$ is determined.
    \item The Koopman mode $\bm{v}_j$ and Koopman eigenvalue $\lambda_j$ in the following equation are determined from $(\bm{y}_0(\bm{R}), \cdots, \bm{y}_n(\bm{R}))$ using dynamic mode decomposition, and the time expansion of $\bm{y}$ is predicted. 
        \begin{eqnarray}
         \bm{y}_{n'}(\bm{R})
          = \sum_{j=1}^{M} (\lambda_j)^{n'} \phi_j(\bm{x}_0)\bm{v}_j  \ \ (n' > n)\nonumber
        \end{eqnarray}
    \item  $\bm{x}_{n'}(\bm{R}) = \bm{g}^{-1}(\bm{y}_{n'}(\bm{R}))$ is determined.
    \item $\bm{x}_{n'}(\bm{R})$ is converted to $\bm{x}_{n'}(\bm{r})$.
        \begin{eqnarray}
        \bm{x}_{n'}(\bm{r}) = \bm{x}_{n'}(\bm{R} - \bm{u}\cdot n') \nonumber
        \end{eqnarray}
\end{enumerate}
\section{Application to short-term precipitation forecasting}
\subsection{Dataset and problem setting}
The present chapter applies the data-driven Koopman operator analysis to short-term precipitation prediction problems. We follow previous research \cite{shi1, shi2} and set up a problem where changes in precipitation are predicted based on time series data of the precipitation distribution from X-band MP radar observations. Through this problem setting, we verify the prediction accuracy of the proposed method.

\begin{figure}[h]
    \centering
    \includegraphics[width=6.8cm]{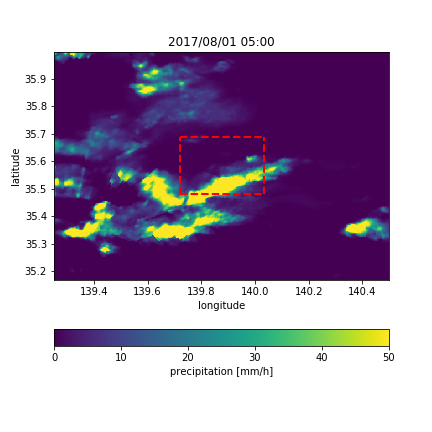}
    \caption{Prediction calculation area of the precipitation distribution, and its assessment subject range (red frame)}
    \label{fig:area}
\end{figure}

This manuscript uses precipitation distribution data observed at 1-min intervals using X-band MP radar. From the prior 30 min of observed data in a given period, we try to predict changes in precipitation 30 min into the future using the proposed method and compare the predicted results to actual precipitation distribution data. Using the computation results for a square area measuring 100 km in size which focuses on Tokyo Bay, as shown in Figure \ref{fig:area}, we set up a square range measuring 25 km in size, which is shown in the red frame, as the assessment subject for prediction accuracy. The observed data used a 250-m mesh, so the computational area had 400 $\times$ 400 meshes, and the prediction subject area had 100 $\times$ 100 meshes.

We used observed data for the 5 h between 05:00 and 10:00 on August 1, 2017, where sufficient precipitation was observed in this range. The precipitation amount 30 min into the future was predicted by conducting the Koopman operator analysis shown in Section 3.2 on the 30 min of observed data from minutes 0 to 29 of each hour, and the result was compared with the measured values from minutes 30 to 59, for a total of five time periods of analyses. The objective of the proposed method was to extract the principal modes while avoiding overfitting of the learning data, so the top five components of the Koopman modes obtained from analysis were used for prediction of the system.

\subsection{Comparative study}
\begin{figure}[h]
    \centering
    \includegraphics[width=6.8cm]{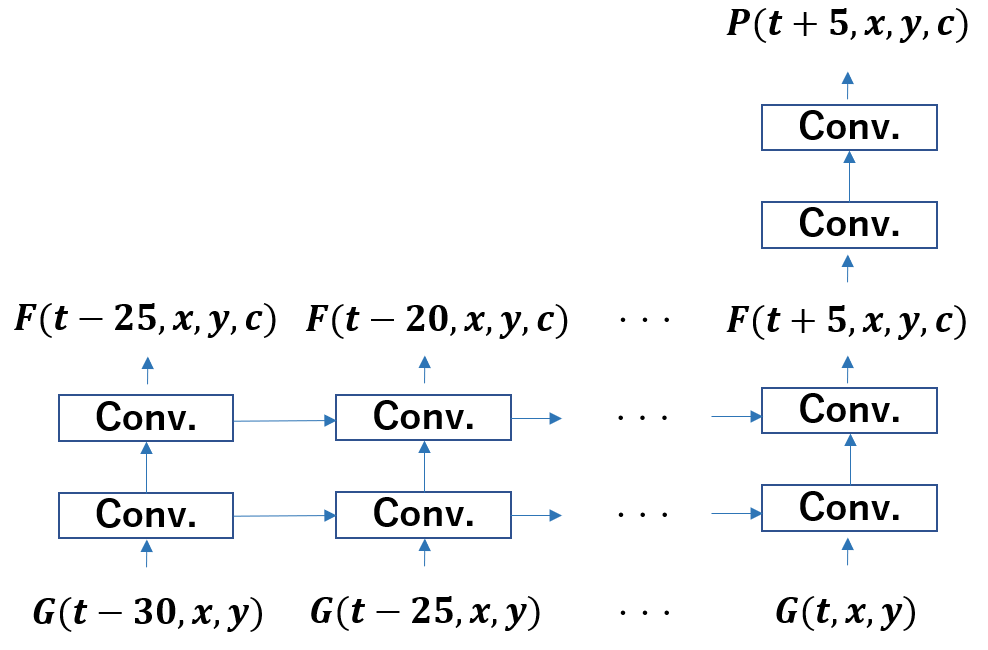}
    \caption{Structure of the deep learning model (Conv-LSTM)}
    \label{fig:architecture}
\end{figure}

The objects of comparison in the precipitation prediction accuracy based on the proposed method included the following: 
\begin{itemize}
    \item Persistent predictions where the precipitation distribution does not change from the current conditions
    \item Kinematically extrapolated predictions where the precipitation distribution in the current conditions is advected owing to the velocity field
    \item Predictions based only on Koopman operator analysis, without kinematic analysis
\end{itemize}
Adding to the three methods listed above, a prediction method based on Conv-LSTM \cite{shi1}, a type of deep learning model, was set up. The model structure is shown in Figure \ref{fig:architecture}, and it predicts precipitation distributions after 5 min of elapsed time using the previous 30 min of precipitation data. $\bm{G}(t, x, y)$ and $\bm{P}(t, x, y)$ in the figure express the measured and predicted precipitation distributions at mesh position $(x,y)$ at time $t$, respectively, and $\bm{F}(t, x, y, c)$ expresses the intermediate feature value in the deep learning model. The channel number $c$ of the characteristic value was set as $30$ ch in the present manuscript. 

This model learned 600 h of precipitation distribution datasets for every month of August from 2011 to 2016 in the subject area, and predictions up to 30 min in advance were conducted by recursive applications on the same data as in the proposed method.

\section{Results and discussions}
The predicted and measured distributions of precipitation with each method at mesh position $(x, y)(1 \le x, y \le  400)$ after $t$ min were set as $P(t, x, y)$ and $G(t, x, y)$, respectively, and the prediction errors were assessed using the normalized mean squared error (NMSE).
\begin{equation}
    NMSE = \frac{\sum_{x,y} (G(t, x, y) - P(t, x, y))^2}{\sum_{x,y} G(t, x, y)^2}
\end{equation}
The total sum at $x,y$ in the above equation was taken as the prediction subject range. The mean squared error of the precipitation prediction results in the prediction subject range was normalized by the measured values for the NMSE.

\begin{figure}[h]
    \centering
    \includegraphics[width=8.0cm]{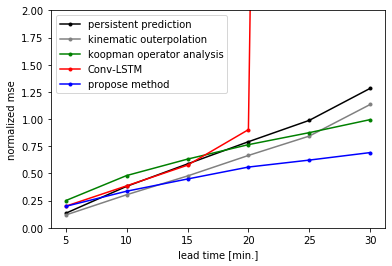}
    \caption{Changes in the precipitation prediction error due to the proposed method}
    \label{fig:pred}
\end{figure}

Figure \ref{fig:pred} shows the average values of the transitions in prediction errors based on each method across the five periods alongside the lead time. The proposed method had the relatively smaller error, especially smallest after 15 min of lead time comparing to all other methods, which showed its effectiveness. The increased error with extended lead time seen in the persistent prediction and the kinematic extrapolation method were controlled in the proposed method, which implies that temporal developments or decay of precipitation, which cannot be expressed with global advection alone, were expressed by the proposed method. Furthermore, the prediction errors in the proposed method were considerably lower than the predictions based only on Koopman operator analysis without kinematic analysis, indicating the effectiveness of applying Koopman operator analysis under Lagrangian coordinates.
Conv-LSTM, a type of data-driven method similar to the proposed method, had an excessively large error after 20 min of lead time, and its prediction failed. In contrast, the proposed method succeeded in stable measurements up to 30 min into the future without exhibiting this type of failure.

With the results corresponding to data at 08:00 on August 1, 2017, we compared the prediction results of the changes in precipitation distribution with the proposed method before and after the hybrid use of kinematic analysis.

\begin{figure*}[hp]
	\centering
	\begin{tabular}{ccc}
	\includegraphics[width=5.0cm]{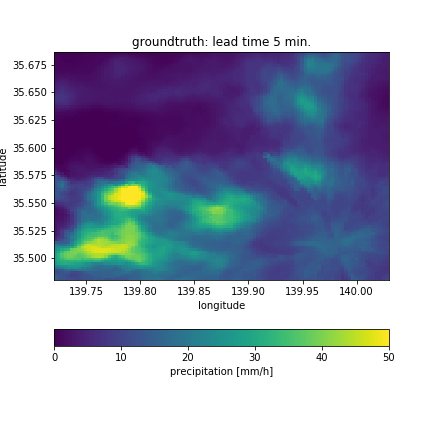} &
	\includegraphics[width=5.0cm]{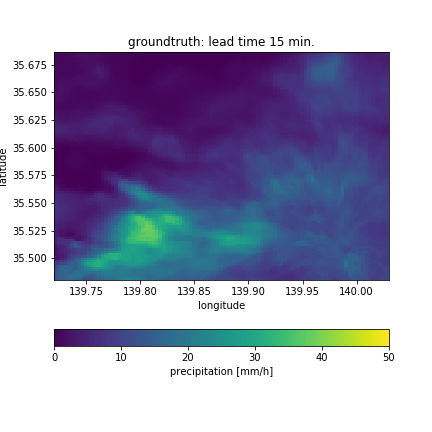} &
	\includegraphics[width=5.0cm]{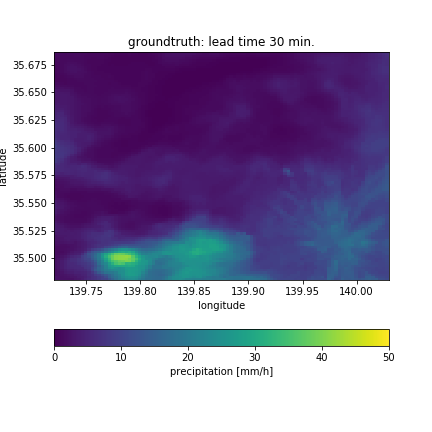} \\
	After 5 min & After 15 min  & After 30 min \\
    \multicolumn{3}{c}{(a)Measured values} \\ \\ \\
    \includegraphics[width=5.0cm]{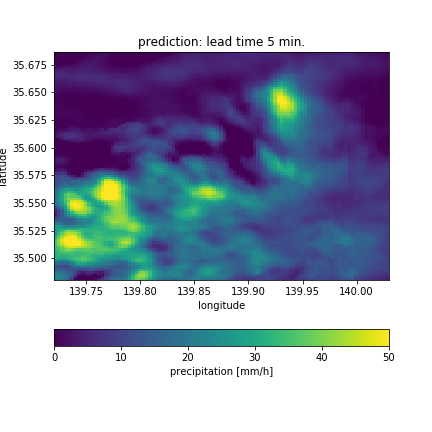} &
	\includegraphics[width=5.0cm]{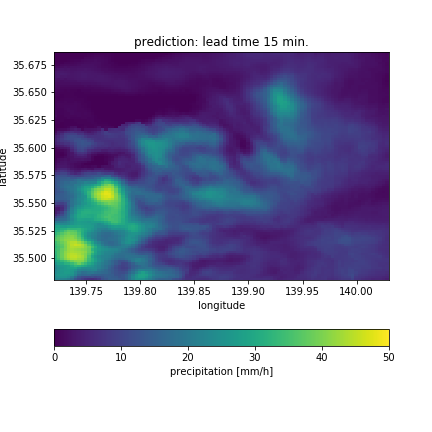} &
	\includegraphics[width=5.0cm]{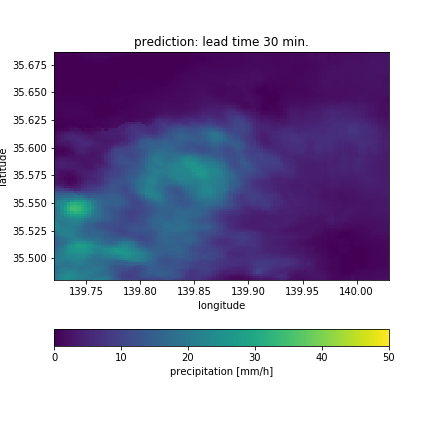} \\
	After 5 min & After 15 min  & After 30 min \\
	\multicolumn{3}{c}{(b)Prediction result: prior to introduction of kinematic analysis }\\  \\ \\
	\includegraphics[width=5.0cm]{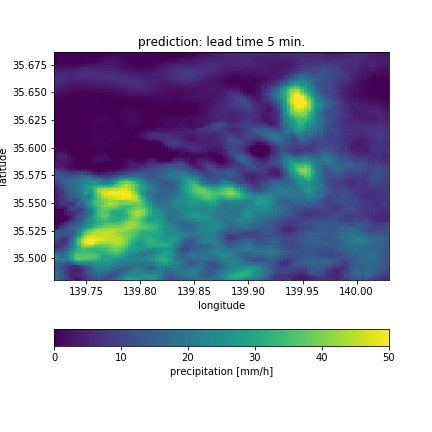} &
	\includegraphics[width=5.0cm]{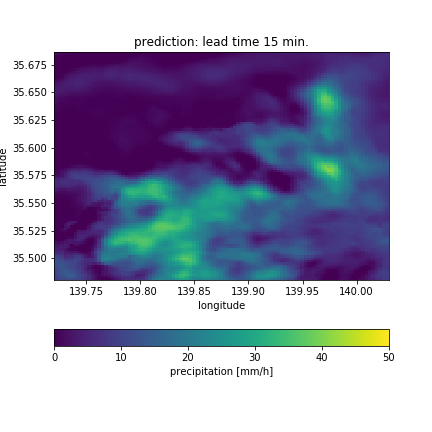} &
	\includegraphics[width=5.0cm]{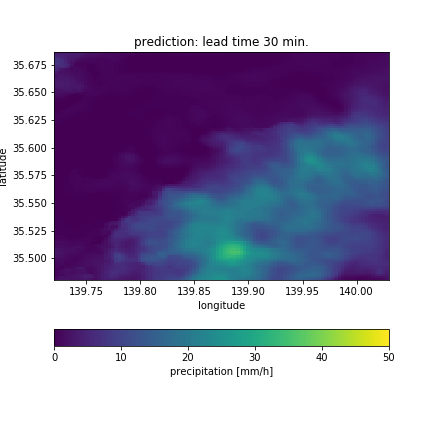} \\
	After 5 min & After 15 min  & After 30 min \\
	\multicolumn{3}{c}{(c)Prediction results: after introduction of kinematic analysis}\\ \\
	\end{tabular}
\caption{Example of precipitation distribution prediction results based on the proposed method and its comparison with measured values}
\label{fig:prediction_comparison}
\end{figure*}

Figure \ref{fig:prediction_comparison} shows the prediction results of changes in precipitation compared with observed values. From the observed values of changes in precipitation shown in (a), we can see that the precipitation in the subject area slowly moved from the prediction starting position to the southeast while steadily decaying and that a precipitation area remained in the southern and eastern sections after 30 min. In contrast, the prediction results shown in (b), the results before the introduction of kinematic analysis, showed that precipitation gradually decayed with time in a manner similar to those seen in the observed values, but the precipitation area was very different. In the prediction results shown in (c), where global advection effects were considered using kinematic analysis, both the decaying tendency of the precipitation and an improvement in the predicted location of the precipitation area were observed. 

Both methods struggled to sufficiently reproduce the complex changes in the precipitation distribution shape seen in the observed values, and the following factors were considered to be the reasons. Essentially, changes in precipitation are phenomena resulting from a variety of physical quantities such as air temperature, air pressure, and specific humidity, and only the precipitation value was used as input data in the prediction model structure. Furthermore, the observable function g in equation (\ref{eqn:kmd}) is an identity function when applying a Koopman operator analysis, but this is not necessarily optimal, and there remains a possibility for setting better observable functions based on the advanced methods mentioned above.

\begin{figure*}[hp]
	\centering
	\begin{tabular}{cc}
	\includegraphics[width=3.8cm]{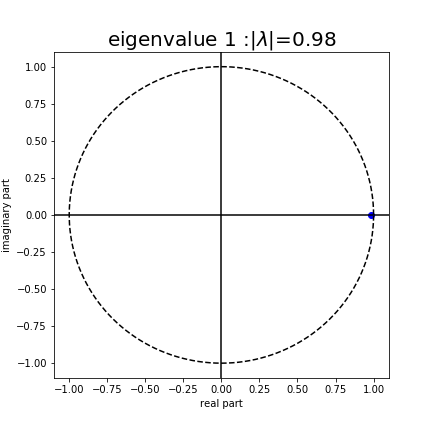} &
	\includegraphics[width=3.8cm]{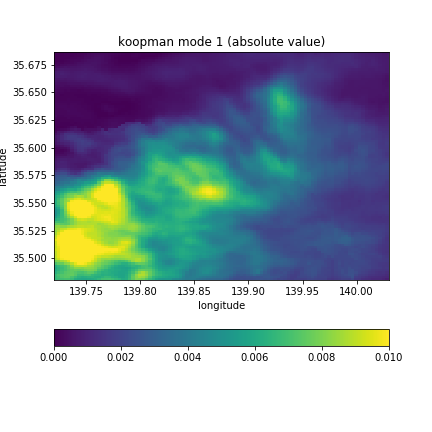}  \\
	 \multicolumn{2}{c}{(a)1st mode}  \\
	\includegraphics[width=3.8cm]{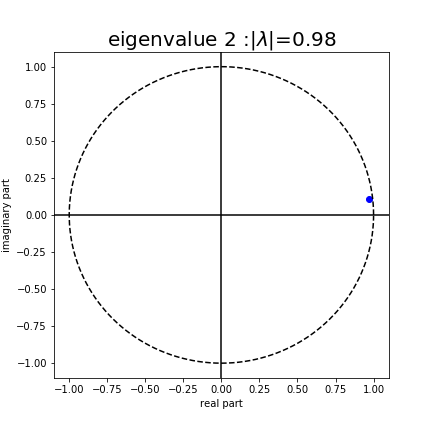} &
	\includegraphics[width=3.8cm]{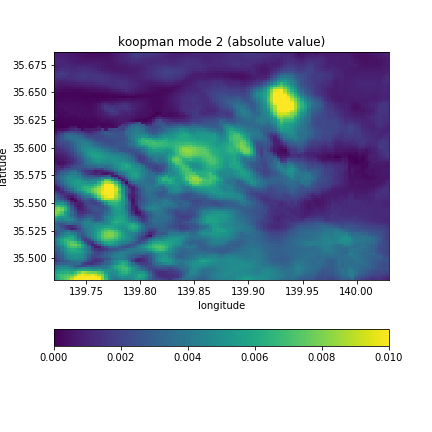}  \\
	 \multicolumn{2}{c}{(b)2nd mode}  \\
	 \includegraphics[width=3.8cm]{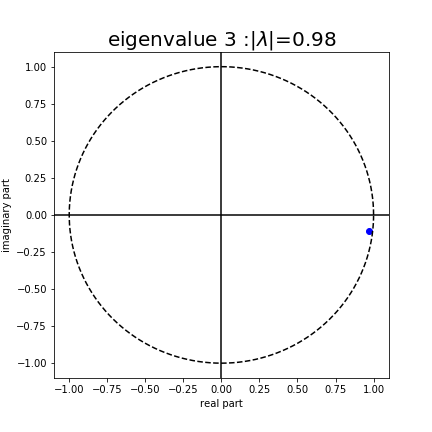} &
     \includegraphics[width=3.8cm]{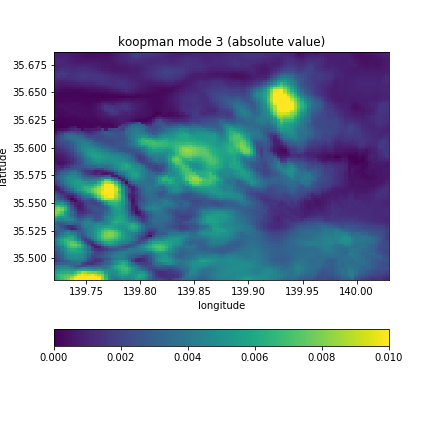}  \\
	 \multicolumn{2}{c}{(c)3rd mode}  \\
	\includegraphics[width=3.8cm]{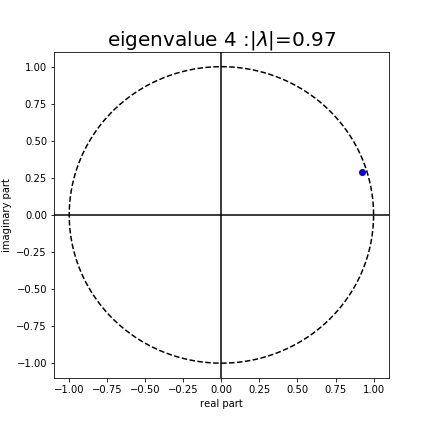} &
	\includegraphics[width=3.8cm]{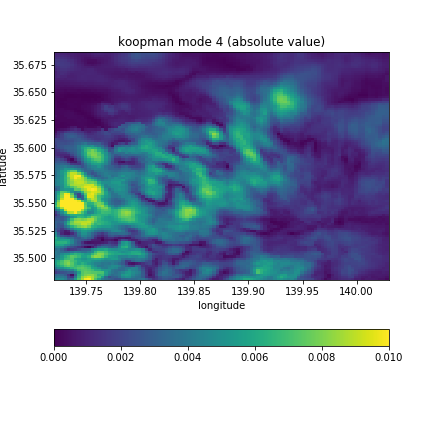}  \\
	 \multicolumn{2}{c}{(d)4th mode}  \\
	\includegraphics[width=3.8cm]{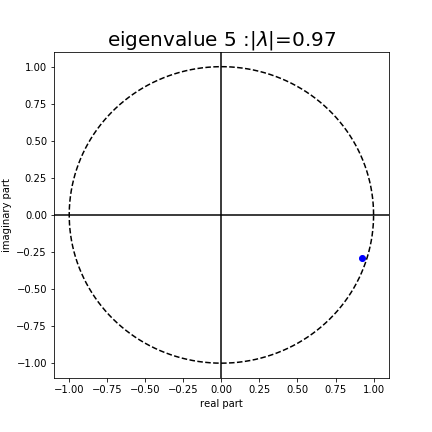} &
	\includegraphics[width=3.8cm]{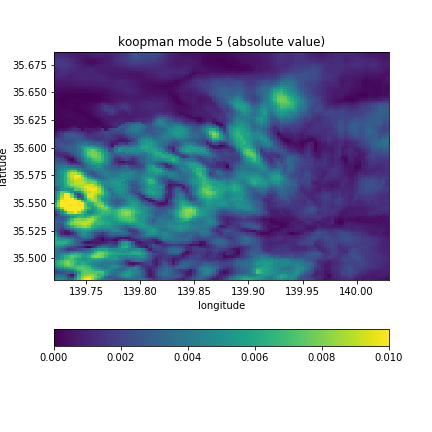}  \\
	 \multicolumn{2}{c}{(e)5th mode} 
	 \end{tabular}
\caption{Top five components of the Koopman mode based on the proposed method (prior to introduction of kinematic analysis), left: Koopman eigenvalue, right: Koopman mode (absolute value)}
\label{modes_noshift}
\end{figure*}

\begin{figure*}[hp]
	\centering
	\begin{tabular}{cc}
	\includegraphics[width=3.8cm]{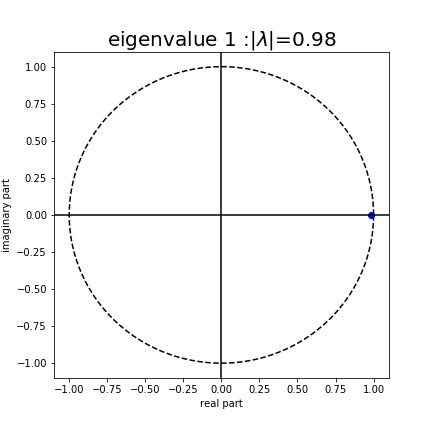} &
	\includegraphics[width=3.8cm]{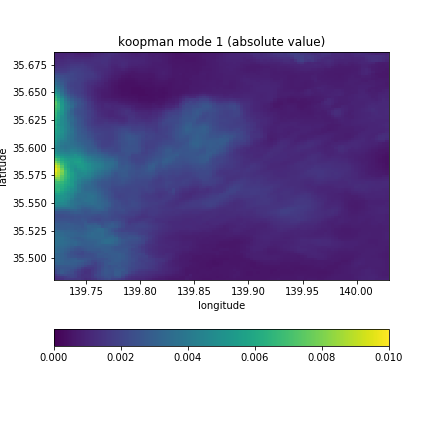}  \\
	 \multicolumn{2}{c}{(a)1st mode}  \\
	\includegraphics[width=3.8cm]{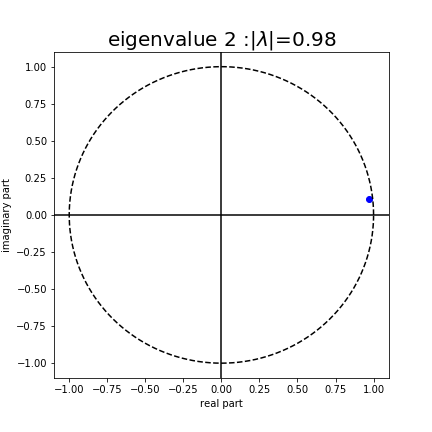} &
	\includegraphics[width=3.8cm]{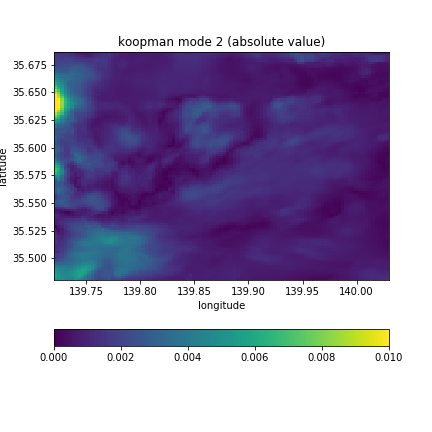}  \\
	 \multicolumn{2}{c}{(b)2nd mode}  \\
	 \includegraphics[width=3.8cm]{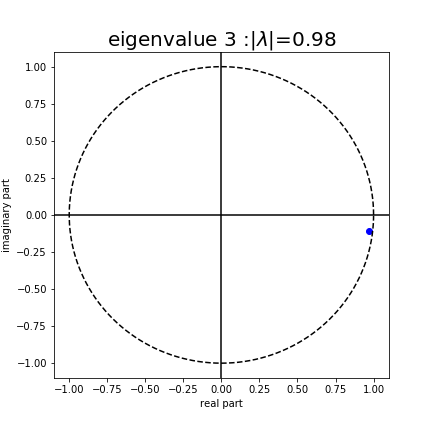} &
     \includegraphics[width=3.8cm]{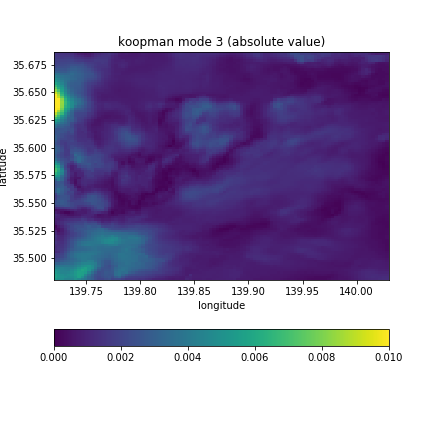}  \\
	 \multicolumn{2}{c}{(c)3rd mode}  \\
	\includegraphics[width=3.8cm]{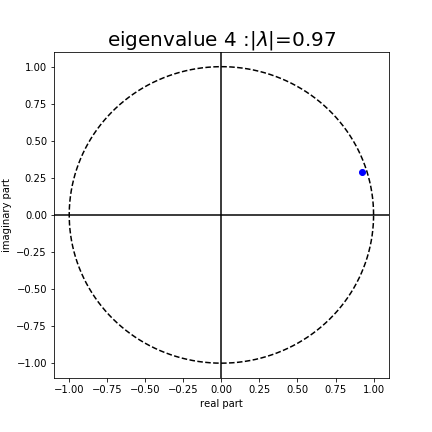} &
	\includegraphics[width=3.8cm]{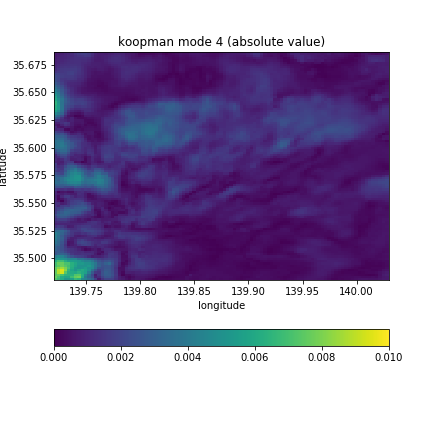}  \\
	 \multicolumn{2}{c}{(d)4th mode}  \\
	\includegraphics[width=3.8cm]{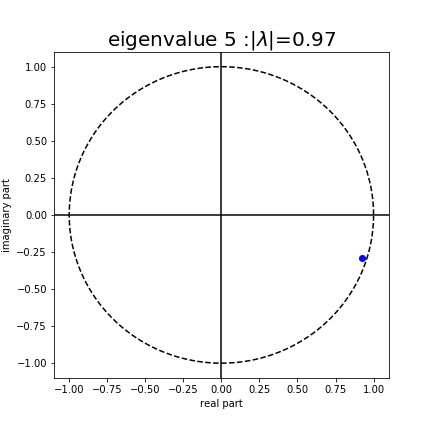} &
	\includegraphics[width=3.8cm]{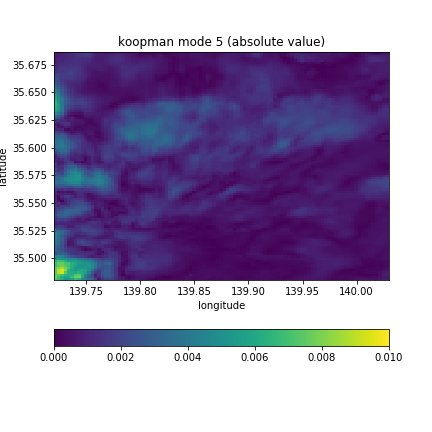}  \\
	 \multicolumn{2}{c}{(e)5th mode} 
	 \end{tabular}
\caption{Top five components of the Koopman mode based on the proposed method (after introduction of kinematic analysis), left: Koopman eigenvalue, right: Koopman mode (absolute value)}
\label{modes_withshift}
\end{figure*}

Finally, the top five components of the Koopman eigenvalue and Koopman mode relating to the changes in the precipitation distribution obtained from the application of the proposed method are shown before and after the introduction of kinematic analysis in Figure \ref{modes_noshift} and \ref{modes_withshift}, respectively. Based on comparisons between the two figures, the absolute value of the eigenvalue of the primary five modes in both results was lower than 1, regardless of the kinematic analysis. This indicates that each mode vibrated while decaying, and the overall precipitation decaying tendency seen in Figure \ref{fig:prediction_comparison}(a) was captured by eigenvalues obtained from Koopman operator analysis.

Furthermore, we can see from comparisons of the Koopman mode shapes in each figure that the results prior to kinematic analysis had spatial mode shapes that expanded throughout the area, whereas the results after kinematic analysis had a mode shape in which a high value was localized, and their characteristics were considerably different. These results occurred because the advected precipitation area was expressed with a spatially fixed mode in fugure \ref{modes_noshift}, whereas the results in Figure \ref{modes_withshift}, where mode decomposition was applied to data whose advection effects were cancelled using kinematic analysis, showed that the mode of the precipitation area that appeared locally in each period was properly handled.

\section{Conclusion}
In order to apply Koopman operator analysis on physical phenomena accompanying advection, the present manuscript decomposed temporal changes in physical quantities between global advection currents and physical changes under Lagrangian coordinates. The advection currents were estimated by kinematic analysis, and the changes in physical quantities were estimated by Koopman operator analysis. The proposed method was applied to actual precipitation distribution data, and the results showed that the development and decay of precipitation were properly captured relative to conventional methods and that stable predictions over long periods were possible.

Future directions of research include the following. First, the present study approximated the velocity field of the atmosphere as uniform and not dependent on time or position, but the accuracy of the proposed method is expected to increase when this approximation is eliminated and detailed velocity field information is predicted from observed data or numerical analyses. Furthermore, the present study used only radar-based precipitation distribution values as input data and kinematically analyzed the influence of the global advection of precipitation distributions. However, changes in precipitation are intrinsically associated with various physical quantities such as air temperature, air pressure, specific humidity, and wind velocity. Thus, incorporating these variables into computational models may be effective at increasing prediction accuracy and discovering the governing equations of these phenomena via Koopman operator analysis. Furthermore, the present research set the observable function used in Koopman operator analysis as an identity function, but we expect improvements in prediction accuracy by the function based on discussions relating to governing equations followed by precipitation.

\section*{Acknowledgement}
The authors would like to express our gratitude for useful comments from Professor Riki Honda in University of Tokyo.

\bibliographystyle{unsrt}  
\bibliography{koopman_kinematics}  

\end{document}